\documentstyle[aps,prl,epsf]{revtex}

\begin{document}
\draft
\title{On the Space Time of a Galaxy}
\author{Tonatiuh Matos and F. Siddhartha Guzm\'{a}n}
\address{Departamento de F\'{\i}sica,\\
Centro de Investigaci\'on y de Estudios Avanzados del IPN,\\
A.P. 14-740, 07000 M\'exico D.F., MEXICO.}
\date{\today}
\maketitle

\begin{abstract}
We present an exact solution of the averaged Einstein's field equations in
the presence of two real scalar fields and a component of dust with
spherical symmetry. We suggest that the space-time found provides the
characteristics required by a galactic model that could explain the
supermassive central object and the dark matter halo at once, since one of
the fields constitutes a central oscillaton surrounded by the dust and the
other scalar field distributes far from the coordinate center and can be
interpreted as a halo. We show the behavior of the rotation curves all along
the background. Thus, the solution could be a first approximation of a
``long exposition photograph'' of a galaxy.
\end{abstract}



\section{introduction}

Doubtless, one of the most interesting open questions in physics by now is
the one concerning the nature of the dark matter in the Universe. In a
series of recent works we have proposed that the dark matter in Cosmos is of
scalar field nature \cite{QSDMCQG,COSPRD,DMCQG,SPHPRD}. Following an
analogous procedure as in particle physics, we wrote a Lagrangian with all
the terms needed to reproduce the observed Universe. In particular, using
the scalar field potential for the scalar field dark matter (SFDM)

\begin{equation}
V(\Phi )=V_{0}\left[ \cosh {(\lambda \sqrt{\kappa_{0}}\Phi )-1}\right]
\label{cosh}
\end{equation}

\noindent where $\kappa _{0}=8\pi /M_{pl}^{2}$, we were able to reproduce
all the successes of the standard Lambda cold dark matter model 
($\Lambda$CDM) at cosmological level very well including
galactic scales \cite{QSDMCQG,COSPRD}. The free parameters of the scalar
potential $V_{0}$ and $\lambda $ can be fitted by cosmological observations
obtaining $\lambda \simeq 20.23$ and $V_{0}\simeq (3\times
10^{-27}\,M_{Pl})^{4}$, being $M_{Pl}$ the Planck mass\cite{COSPRD}. The
mass of the scalar particle is then $m_{\Phi }\simeq 9.1\times
10^{-52}M_{Pl}=1.1\times 10^{-23}$eV (compare this value of the scalar mass
with that in \cite{Hu}) . Under galactic scales there are some differences
between the $\Lambda $CDM\ and the SFDM. The self interaction of the scalar
field $\Phi $ could explain the observed dearth of dwarf galaxies and the
smoothness of the galaxy core halos as well \cite{COSPRD}. The question
whether this scalar field is from fundamental origin or it is a combination
of some other fundamental particles, is open. Nevertheless, in order to
reproduce the high resolution N-body numerical simulations of self
interacting dark matter\cite{vladimir}, we found that the renormalization
scale of this scalar field is of the same order of the Planck mass \cite
{CROSS}. This would suggest that this scalar field dark matter could have a
fundamental origin. We have also studied the scalar field hypothesis at
galactic level in \cite{DMCQG,SPHPRD}. The idea follows the standard idea of
galaxy formation, namely that scalar field (dark matter) fluctuations are
the responsible for the origin of the galaxies. In the case of the scalar
field potential (1), we have $\cosh (\lambda \Phi )\rightarrow \cosh
(\lambda (\Phi +\delta \Phi ))\sim \exp (\lambda \delta \Phi )$ for regions
where the scalar field fluctuation dominates $\delta \Phi >\Phi $. Of
course, as in the standard theory of galaxy formation, the dark matter
fluctuations are of different size in different regions of the Universe, for
different galaxies. Therefore, at galactic level, we have a scalar field
potential which depends on local variables. Thus, the exponential potential
approximates the $\cosh $ potential in some regimes of the scalar field and
we could develop some interesting aspects of a scalar dark matter halo in
galaxies.\newline

On the other hand, new observations show that supermassive central objects
lying in active galactic nuclei seem to be correlated with the velocity
dispersion of the dark matter composing the dark halo, suggesting that the
central object was formed at the same time than the halo\cite{apj}. This is
probably contrary to the standard idea about galactic nuclei that proposes
the existence of a central black hole, since it should be formed a time
after the disc, $i.e.$ much more after the formation of the halo. Going
further, it has been also shown that the supermassive objects which are
supposed to be in the center of galactic nuclei, could be boson stars,
obtaining the same predictions for the rate of accretion of matter due to
the presence of a completely regular space-time background without surface
or horizon\cite{diego}.\newline

The question we are facing now is whether there is a dynamical mechanism
that could provide a realistic scenario of galaxy formation using the scalar
field dark matter hypothesis. First of all, a complete evolution of galactic
and under galactic fluctuations belong to the non-linear regime of
perturbations. The right answer would be provided by numerical evolution of
Einstein's equations. Fortunately, a partial answer is given in numerical
research on Einstein's equations developed since 1990. In particular, the
collapse of a scalar field has been studied deeply in \cite
{seidel91,seidel94,luis}, and it was found that there are final equilibrium
and stable configurations for collapsed scalar field particles: boson stars
(when the scalar field is complex) and oscillatons (when the scalar field is
real and time-dependent), both of them being formed through a process called
gravitational cooling\cite{seidel94}. Based on these ideas, we present in
section \ref{sec:toymodel} the motivations inviting us to build a galactic
model with scalar fields, followed by the solution to the model in section 
\ref{sec:solution}; physical features of the model are shown in section \ref
{sec:physics} and finally we draw some conclusions.\newline

\section{The Galactic Model with Scalar Field Dark Matter}

\label{sec:toymodel}

We first recall the main results obtained in\cite{seidel91,seidel94}.
Through the gravitational cooling process, a cosmological fluctuation of
scalar field would collapse to form a compact oscillaton by ejecting part of
the scalar field. This ejected part of the cooling process would carry the
excess of kinetic energy out and can play the role of the halo of the
collapsed object. The final configuration then consists of a central compact
object, an oscillaton, surrounded by a diffuse cloud of scalar field, both
formed at the same time due to the same collapse process. This would provide
a correlation between the central object and the scalar halo.\newline

If the central object is an oscillaton, its formation is due to coherent
scalar oscillations around the minimum of the scalar potential~(\ref{cosh}).
The scalar field $\Phi $ and the metric coefficients (considering the
spherically symmetric case) are time dependent and it was proved that this
configuration is stable, non-singular and asymptotically flat\cite{seidel91}%
. But, this could not be the final answer because, from the
quantum-mechanical point of view, a Bose-Einstein condensate should form and
we must take into account that the scalar field $\Phi $ can also decay or
self annihilate\cite{tkachev,jeremy}. In such a case, we should have to
consider the scattering cross section $\sigma _{\Phi \rightarrow \Phi }$ in
order to calculate the relaxation time of such condensation. In the models
treated in\cite{tkachev,jeremy,peebles}, the coefficient of $\Phi ^{4}$ is
the responsible both for the size of the compact object and for the
scattering cross section $\sigma _{\Phi \rightarrow \Phi }.$ With potential~(%
\ref{cosh}) it is necessary to take into account all the couplings\cite
{CROSS}. Nevertheless, it is possible to find the relaxation time smaller
than the age of the Universe if \cite{tkachev}

\begin{equation}
g>10^{-15}(m_{\Phi }/eV)^{7/2}
\end{equation}

\noindent with $g$ being the effective $\Phi ^{4}$-theory coupling. Then,
considering the value $m_{\Phi }\simeq 10^{-23}$ eV (see \cite{COSPRD}), $%
g>10^{-96}$. This condition is well satisfied for our case since $g\sim
10^{-48}$\cite{CROSS}. This ensures that the relaxation time is shorter than
the age of the Universe. \newline

In the case studied in\cite{seidel94}, a massive scalar field without
self-interaction collapses ejecting out 13\% of the initial configuration of
scalar field. The critical mass of the final configuration for the
oscillaton depends on the mass of the boson\cite{seidel91}. For our model
(1), the mass of the scalar field is $m=1.1\times 10^{-23}eV$ and it is
found that

\begin{equation}
M_{crit} \sim 0.6 \frac{M_{pl}^{2}}{m} \sim 10^{12} M_{\odot}  \label{masa}
\end{equation}

\noindent which is a surprising result: the critical mass needed for the
scalar field to collapse is of the same order of magnitude than the dark
matter contents of a standard galaxy. Then, we expect that the fluctuations
of this scalar field due to Jeans instabilities will in general collapse to
form objects with mass of the same order than the mass of the halo of a
typical galaxy. This is the reason by means of which we expect that the SFDM
model can also work at galactic level. Summarizing, we consider two working
hypothesis up to now. First, we identify the formation of a central compact
object and a halo with the gravitational collapse of a scalar field. The
compact object could be an oscillaton (since we are dealing only with a real
scalar field) or a Bose-Einstein condensate. Second, we identify the ejected
scalar field with the halo of this galaxy.\newline

In a realistic model the metric and fields should depend on time and a
complete study would involve numerical calculations within and beyond
General Relativity. One alternative is to study the behavior of the galaxy
numerically with all the hypotheses stated above. This procedure has the
advantage that we do not need to eliminate any {\it a priori} consideration,
but we can loose some important physical information inside of the numbers
we obtain. Other alternative is to perform some approximations on the metric
and fields and find exact solutions. These approximations could also lead us
to loose crucial information of the system but we can keep physics under
control. In our opinion, both procedures must be developed in order to have
a better understanding of the hypothesis.\newline

In this work we will adopt the second alternative, i.e. we will build a toy
model for case stated above in purely geometrical terms and considering only
the final stage of the collapse. We let for a future work the dynamical
evolution\cite{futuro}. To start with, we support our toy model on the
numerical results studied in \cite{seidel91,seidel94,luis}. First, since the
time-dependence of the metric in an oscillaton is quite small \cite{seidel91}%
, we suppose then that the center of this toy galaxy is an oscillaton which
oscillates coherently but considering a {\it static} metric. This is an
approximation because neither the galactic nuclei nor the oscillaton are
expected to be static. However, for the purposes of this analytic work, we
suppose that the dynamics of the oscillaton can be frozen in time in a way
we explain below. Second, we do not expect the scalar halo to possess the
same properties than the collapsed oscillaton; in some sense, they must be
different. Thus, we will consider the scalar halo as another scalar field.
Third, baryonic matter is considered to lie in the galaxy. We suppose that
only the  baryonic matter at the galaxy center and the bulge will contribute
essentially to the curvature of the space-time of the galaxy. This baryonic
component is assumed to be dust. As in previous works \cite{DMCQG,SPHPRD},
we let the luminous matter around the galaxy as test particles, i.e. they do
not essentially contribute to the curvature of the space-time.\newline

\section{The Analytical Solution}

\label{sec:solution}

The Lagrangian density we have to start with reads ${\cal L}=R+ {\cal L}%
_{SFO}+{\cal L}_{SFH}+{\cal L}_{dust}$, being $R$ the scalar curvature and
the former terms correspond to the Lagrangian of the scalar field oscillaton
(SFO), the scalar field halo (SFH) and dust, which should play the role 
of a galactic bulge. For the scalar objects, the
following expressions are available

\begin{equation}
{\cal L}_{SFO}=-\frac{1}{2}(\partial _{\mu }\Phi \partial ^{\mu }\Phi
)-V(\Phi ),~~~~~ {\cal L}_{SFH}=-\frac{1}{2}(\partial _{\mu }\psi \partial
^{\mu }\psi )-U(\psi)
\end{equation}

\noindent Here $\Phi$ is the scalar field corresponding to the oscillaton
and $V(\Phi )$ its potential of self-interaction. $\psi $ is the scalar
field describing the scalar field halo and $U(\psi)$ its respective scalar
potential. As a consequence of the non-coupling of the fields in the total
Lagrangian, the Einstein's equations are written $G_{\mu \nu }=\kappa
_{0}[T_{\mu \nu }^{SFO}+T_{\mu \nu }^{SFH}+T_{\mu \nu }^{dust}]$, with

\begin{eqnarray}
T_{\mu \nu }^{SFO} &=&\partial _{\mu }\Phi \partial _{\nu }\Phi -1/2g_{\mu
\nu }\left[ \partial ^{\sigma }\Phi \partial _{\sigma }\Phi +2 V(\Phi )%
\right]  \nonumber \\
T_{\mu \nu }^{SFH} &=&\partial _{\mu }\psi \partial _{\nu }\psi -1/2g_{\mu
\nu}[\partial ^{\sigma }\psi \partial _{\sigma} \psi +2U(\psi)]  \nonumber \\
T_{\mu \nu }^{dust} &=&du_{\mu }u_{\nu }  \nonumber
\end{eqnarray}

\noindent being $d$ the density of the dust and $u^{\alpha }$ the four
velocity of its particles. Since the scalar field $\Phi $ is time-dependent
oscillating coherently around the minimum of its scalar potential, we can
write $\Phi =P(r)\cos (\omega t)$ (see \cite{luis}). In order to handle
Einstein's equations we average them during the period of a scalar
oscillation, that is, we take $<G_{\mu \nu }=\kappa _{0}[T_{\mu \nu
}^{SFO}+T_{\mu \nu }^{SFH}+T_{\mu \nu }^{dust}]>$. This procedure gives the
lowest order approximation for an oscillaton and we are left with
time-independent differential equations\cite{luis}; in the notation 
of reference \cite{luis} where the metric functions are written as 
$g(r,t)=g_0(r) + g_1(r) \cos(\omega t) + ...$ 
we are looking only for $g_0$ terms of the metric and fields, which are the
dominant ones. Besides of Einstein's equations there are two Klein-Gordon
(KG) equations for the scalar fields,

\begin{equation}
\Phi ^{;\mu }{}_{;\mu }+\frac{dV}{d\Phi }=0,~~~\psi ^{;\mu }{}_{;\mu }+\frac{%
dU}{d\psi }=0
\end{equation}

\noindent The KG on the left depends on time, but it can be shown that one
can factorize an overall cosine-term and then the resulting differential
equation is time-independent\cite{luis}.\newline

Then, starting from a spherically symmetric space-time, using the {\it %
harmonic maps ansatz }we were able to find a solution of the system. In few
words, the main idea behind the harmonic maps ansatz is the
reparametrization of the metric functions with convenient auxiliary
functions which will obey a generalization of the Laplace equation, along
with some consistency relationships; the latter are usually quite difficult
to fulfill, and great care and intuition must be taken in order to get a
system of equations both workable with and interesting enough \cite{ma}. The
exact solution of the averaged Einstein's field equations is

\begin{eqnarray}
\psi &=&\sqrt{\frac{v_{a}^{2}}{2\kappa _{0}}}\ln (r^{2}+b^{2})+\psi _{0} 
\nonumber \\
U(\psi ) &=&\frac{2v_{a}^{2}}{\left( 1-v_{a}^{2}\right) \kappa _{0}}\exp
\left( -\sqrt{\frac{2\kappa _{0}}{v_{a}^{2}}}(\psi -\psi _{0})\right)
\end{eqnarray}

\noindent for the scalar field $\psi$. The scalar field $\Phi $ for the
oscillaton we obtained is

\begin{eqnarray}
V(\Phi (r)) &=&-1/4\,{\frac{{\omega }^{2}\left( {r}^{2}+{b}^{2}\right) ^{-{v}%
_{a}^{2}}rP^{2}}{\left( -r+2\,M\right) B_{0}}}+1/2\,{\frac{\left( 1+{v}%
_{a}^{2}\right) {r}^{2}+{b}^{2}\left( 1-v_{a}^{2}\right) }{\left( {r}^{2}+{b}%
^{2}\right) \kappa _{0}\,{r}^{2}\left( 1-{v}_{a}^{2}\right) }}  \nonumber \\
&&+1/2\,{\frac{-\left( {v}_{a}^{2}+1\right) ^{2}{r}^{4}+M{v}_{a}^{2}\left(
1+2\,{v}_{a}^{2}\right) {r}^{3}-2\,{b}^{2}\left( 1+2\,{v}_{a}^{2}\right) {r}%
^{2}+5\,{v}_{a}^{2}M{b}^{2}r-{b}^{4}}{\kappa _{0}\,{r}^{2}\left( 1-{v}%
_{a}^{4}\right) \left( {r}^{2}+{b}^{2}\right) ^{2}}}
\end{eqnarray}

\noindent where the function $P$ is given up to quadratures

\begin{equation}
P=\int \sqrt{-2\,{\frac{\left( 1-{v}_{a}^{4}\right) }{\left( r-2\,M\right)
\kappa _{0}\,r}}+2\,{\frac{\left( 1-{v}_{a}^{4}\right) {r}^{4}-{v}%
_{a}^{2}M\left( 3-2\,{v}_{a}^{2}\right) {r}^{3}+2\,{r}^{2}{b}^{2}-3\,{v}%
_{a}^{2}M{b}^{2}r+{b}^{4}}{\left( {r}^{2}+{b}^{2}\right) ^{2}\left(
r-2\,M\right) \kappa _{0}\,r}}}{dr}
\end{equation}

\noindent being $v_{a}$ an asymptotic value of the tangential velocity of
the test particles in our toy galaxy. The density of the dust is given by

\begin{eqnarray}
d &=&{\frac{1}{\kappa_0{r}^{2}}-}1/2\,{\frac{{\omega }^{2}\left( {r}^{2}+{b}%
^{2}\right) ^{-{v}_{a}^{2}}rP^{2}}{B_0\left(r-2\,M\right)}}  \nonumber \\
&&-{\frac{\left( 1-{v}_{a}^{4}\right) {r}^{4}-{v}_{a}^{2}M\left( 3-2\,{v}%
_{a}^{2}\right) {r}^{3}+2\,{b}^{2}\left( 1-{v}_{a}^{2}\right) {r}^{2}+{v}%
_{a}^{2}M{b}^{2}r+{b}^{4}}{\kappa _{0}\,{r}^{2}\left( 1-{v}_{a}^{4}\right)
\,\left( {r}^{2}+{b}^{2}\right) ^{2}}}
\end{eqnarray}

\noindent Finally, the corresponding line element reads

\begin{equation}
ds^{2}=-B_{0}(r^{2}+b^{2})^{v_{a}^{2}}\left( 1-\frac{2M}{r}\right) dt^{2}+%
\frac{A_{0}}{(1-\frac{2M}{r})}dr^{2}+r^{2}d\Omega ^{2}  \label{met_bosones}
\end{equation}

\noindent with $d\Omega ^{2}=d\theta ^{2}+\sin ^{2}\theta d\varphi ^{2}$ and 
$M$ is a constant with the interpretation discussed below. This metric is
singular at $r=0$, but it has an event horizon at $r=2M$. This metric does
not represent a black hole because it is not asymptotically flat.
Nevertheless, for regions where $r\ll b$ but $r>2M$\ the metric behaves like
a Schwarzschild black hole. Inside of the horizon the pressure of the
perfect fluid is not zero anymore, thus our toy model is valid only in
regions outside of the horizon, where it could be an approximation of the
galaxy.\ Metric (\ref{met_bosones}) is not asymptotically flat, but it has a
natural cut off when the dark matter density equals the intergalactic
density as mentioned in \cite{SPHPRD}.\newline

\section{Physical Features of the Model}

\label{sec:physics}The dust density and the oscillaton scalar field
potential depend on the value of the function $P$. We find that $P$ goes
very rapidly to a positive constant value, which depends on the boundary
conditions one imposes. Let its asymptotic value be $P_{0}$ when $r \gg 2M$
whose interpretation would be the asymptotic amplitude of $\psi $.\newline


In order to understand the other parameters of the metric, let us proceed in
the following way. It is believed that in a standard galaxy, the central
object has a mass of $M\sim 2-3\times 10^{6}M_{\odot }\sim $some $a.u.$ Far
away from the center of the galaxy,\ say from $1pc$ up, the term $2M/r\ll 1$%
. In this limit metric (\ref{met_bosones}) becomes

\begin{equation}
ds^{2}=-B_{0}(r^{2}+b^{2})^{v_{a}^{2}}dt^{2}+A_{0}dr^{2}+r^{2}d\Omega ^{2}.
\label{met_sph}
\end{equation}

\noindent This space-time is very similar to metric (18) of reference \cite
{SPHPRD}, but now with the potential

\begin{equation}
U(\psi (r))=\frac{2v_{a}^{2}}{\kappa _{0}\left( 1-v_{a}^{2}\right) }\frac{1}{%
(r^{2}+b^{2})}
\end{equation}

\noindent being both solutions the same in the limit $r\rightarrow \infty $.
This implies that $A_{0}=1-v_{a}^{4}$, recovering in this way the asymptotic
results shown in \cite{SPHPRD}. Parameter $b$ is related to parameter $b$ of
metric (21) in reference \cite{DMCQG}, where it acts as a gauge parameter.
Of course this metric is only valid far away from the center of the galaxy.
With parameter $b$ it is now possible to fit quite well the rotation curves
of spiral galaxies. Therefore metric (\ref{met_bosones}) could not only
represent the exterior part of the galaxy, but it could be a good
approximation for the core part of it as well. Let us see this point.\newline

The rotation curves $v^{rot}$ seen by an observer at infinity for a
spherically symmetric metric are given by $v^{rot}=\sqrt{%
rg_{tt}{}_{,r}/(2g_{tt})}$ \cite{SPHPRD}. For metric (\ref{met_bosones})
such result reads

\[
v^{rot}(r)= \sqrt{\frac{v^{2}_{a}(r-2M)r^2 +M(r^2+b^{2})}{(r-2M)(r^2+b^{2})}}
\]


\noindent formula that allows one to fit observational curves. In Figure 1
it is shown the change of the rotation curve when the value of $M$ changes,
and it is evident that such fact would affect only the kinematics in the
central parts of the galaxy, exactly in the same way as the mass of the
central object should do \cite{diego}.\newline

\begin{figure*} 
\centerline{ 
\epsfxsize=5cm \epsfbox{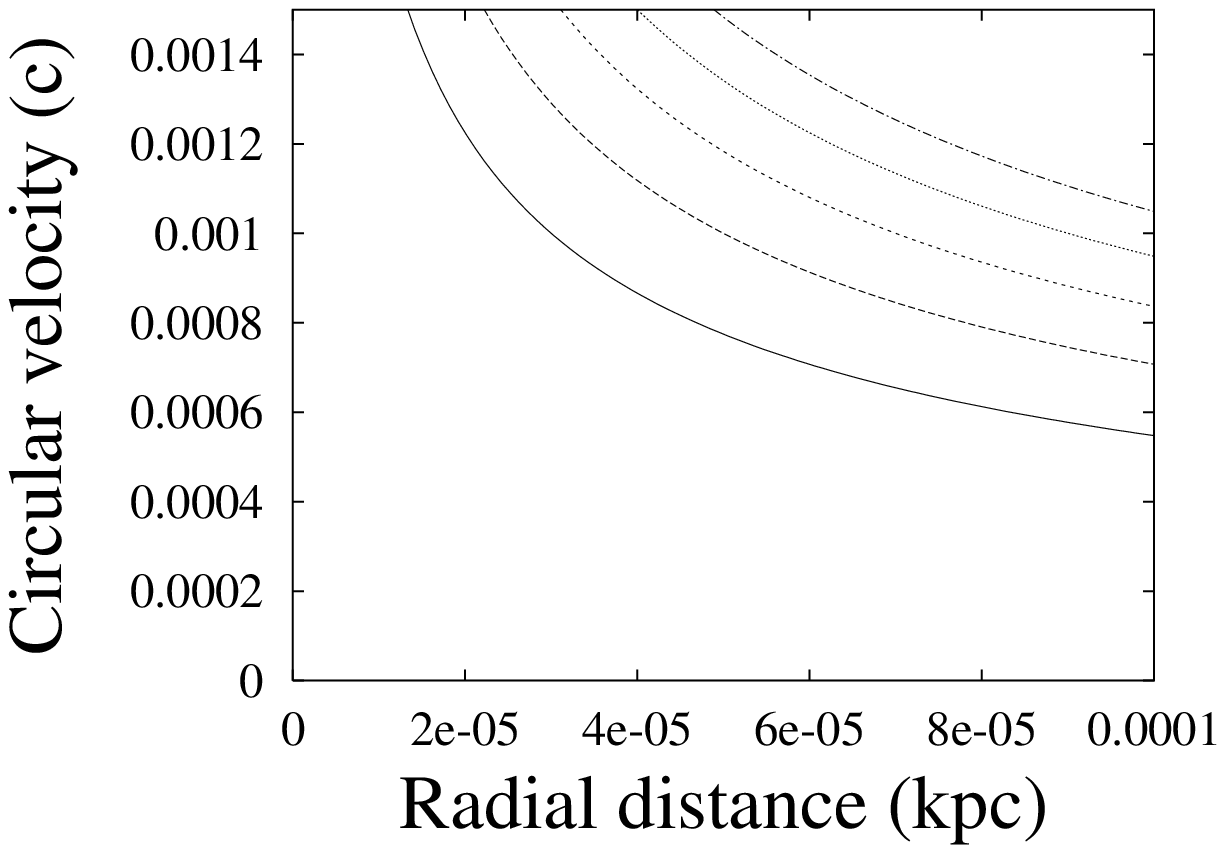} \epsfxsize=5cm \epsfbox{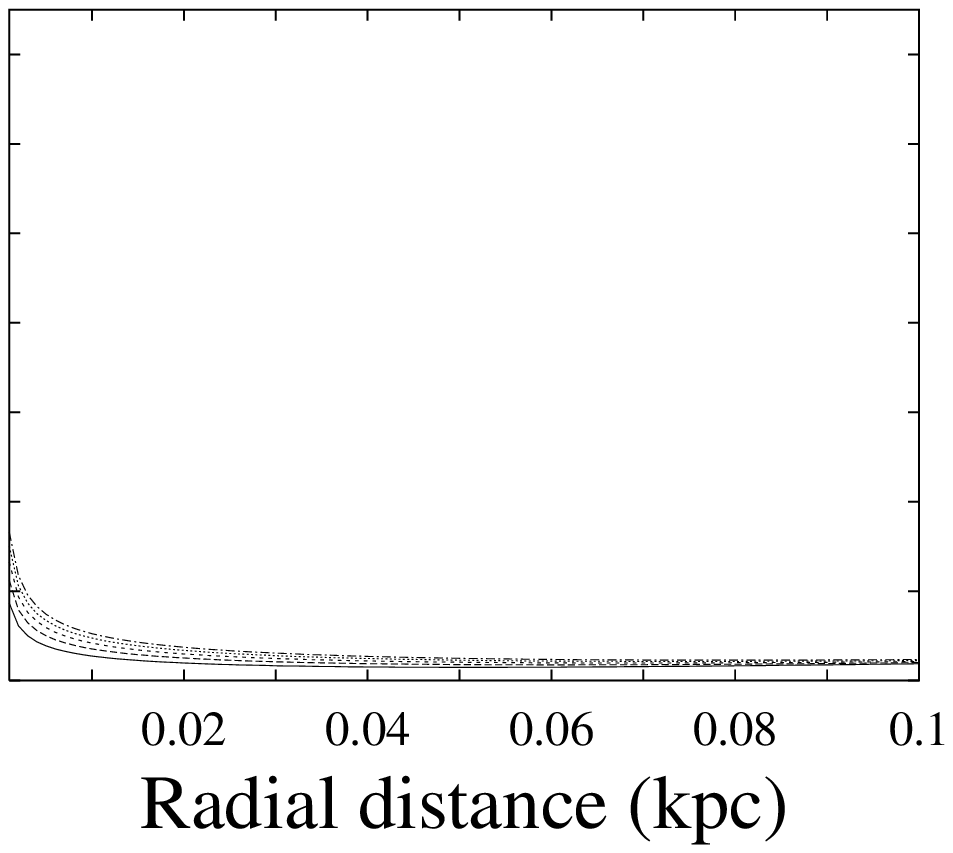}
\epsfxsize=5cm \epsfbox{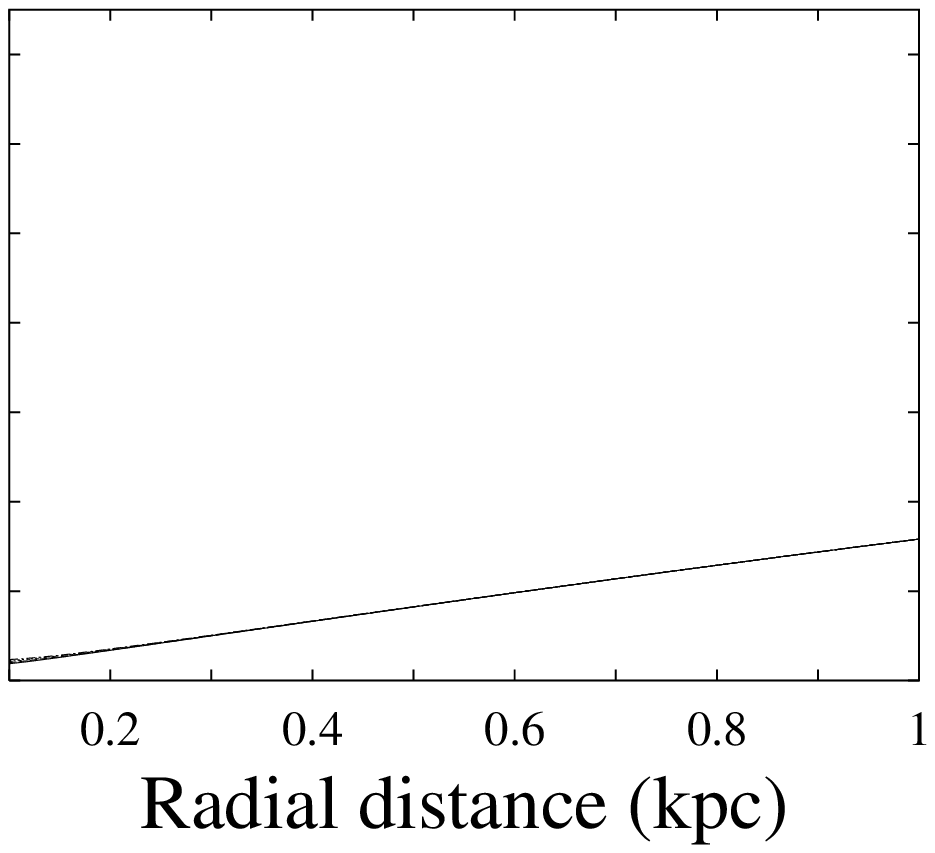} \epsfxsize=5cm \epsfbox{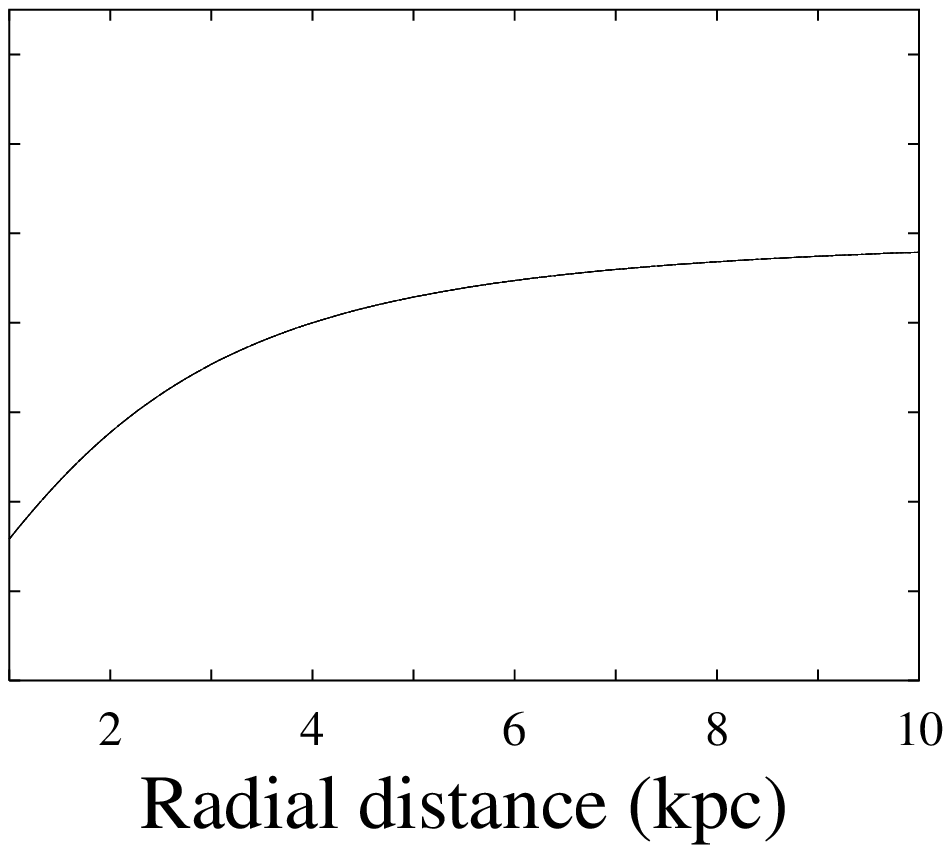}
}~\vspace{0.5cm}
\caption{Rotation curves provided by the line element (\ref{met_bosones}),
for different values of the mass of the central object: $M=3,5,7,9,11\times
10^{6}M_{\odot }$. In the left side it is shown the rotation curve in the
innermost part of the galactic nuclei, in the middle an intermediate radial
distance an finally in the right hand side it is shown the properly called
rotation curve associated to the dark matter.}
\label{fig:fig1}
\end{figure*}

Nevertheless, it calls the attention in Figure 1 that there is a region
where the velocity of test particles is near to zero, and in fact the
rotation curve decays in a Keplerian way. Let us see what happens in such a
region near the galactic center: the factor $(r^{2}+b^{2})^{v_{a}^{2}}$ in $%
g_{tt}$ is almost 1 for $r\sim 10pc$ for reasonable values $b\sim $ $kpc$
and $v_{a}\sim 10^{-3}$, therefore the time-like geodesics are determined by
the factor $1-2M/r$. Under such condition the contribution of $\psi $ is
accurately zero for an observer far away from the galaxy. The interesting
situation comes after calculating the density of all the components to the
central density, it reads

\begin{eqnarray}
\rho _{dust}+\rho _{SFO}+\rho _{SFH} &=&-1/2\,{\frac{M\left( r-2\,M\right)
\left( {r}^{2}+{b}^{2}\right) ^{{v}_{a}^{2}-2}B_{0}\,}{{r}^{6}\kappa
_{0}\,\left( 1-{v}_{a}^{4}\right) }}\left( -4\,{r}^{6}{v}_{a}^{2}\right. 
\nonumber \\
&&\left. +{b}^{2}\left( 4-7\,{v}_{a}^{2}+2\,{v}_{a}^{4}\right) {r}^{4}+3\,{b}%
^{4}\left( 3-{v}_{a}^{2}\right) {r}^{2}+3\,{b}^{6}\right)  \label{trash}
\end{eqnarray}


\noindent where the approximation $v_{a}\ll 1$ and $%
(r^{2}+b^{2})^{v_{a}^{2}}\sim 1$ is considered to be valid for reasonable
values of the parameters. Moreover, we can infer that the total energy
density as seen by an observer far away from the center of the galaxy
diverges when $r$ approaches to zero and goes rapidly to zero as $1/r^{5}$
when $r\rightarrow \infty $. But an observer at infinity measures the mass
of the central object, it has not the possibility to distinguish between the
density of each component.\newline

Now let us explore an ADM-like concept of mass associated to the central
region; we stress that metric (\ref{met_bosones}) is not asymptotically
flat and therefore the $ADM$ concept of mass strictly fails to be valid, so 
we recall that there is a region into the interval $2M \ll r < b$
geometrically almost flat, where we will define a useful infinity
$\infty_f$ that should serve to calculate the mass of the central
configuration through the standard metric

\begin{equation}
ds^{2}=-e^{2\delta }dt^{2}+\frac{dr^{2}}{(1-\frac{2m}{r})}+(1-\alpha
)r^{2}d\Omega ^{2}  \label{estand}
\end{equation}

\noindent where $m=m(r)$ is interpreted as the mass function and $\delta
=\delta (r)$ as the gravitational potential. This form of the metric is
convenient because in this coordinates $m_{,r}=4\pi r^{2}\rho _{T}$, where $%
\rho _{T}$ is the total density of the object. This interpretation is
correct in regions where the space-time is almost flat, i.e. far away from
the horizon $r=2M$. Close to the horizon or inside of it, function $m$ is a
quantity that should be similar to the mass of the object, but it is not
since it contains the contribution of all the components together; in this
region, where the curvature of the space-time is huge, the volume element is
different from $4\pi r^{2}dr$. Furthermore, inside of the horizon we are not
able to know the real physics of the object. On the other side, far away
from the center of the toy galaxy, this function can be interpreted as the
mass of an infinitesimal shell at radius $r$. Anyway we will call function 
$m$ the mass function everywhere. Thus, the ADM-like mass is obtained
by $M_{ADM}=%
\mathrel{\mathop{\lim }\limits_{r\rightarrow \infty_f }}%
m$. We perform the coordinate transformation $\sqrt{A_{0}}r\rightarrow r$, $%
\sqrt{A_{0}}b\rightarrow b$\ in order to compare metrics (\ref{met_bosones})
and (\ref{estand}). We obtain

\begin{equation}
ds^{2}=-\frac{B_{0}}{A_{0}^{^{v_{a}^{2}}}}\left( r^{2}+b^{2}\right)
{}^{v_{a}^{2}}\left( 1-\frac{2M\sqrt{A_{0}}}{r}\right) dt^{2}+\frac{dr^{2}}{%
(1-2M\sqrt{A_{0}}/r)}+\frac{r^{2}}{A_{0}}d\Omega ^{2}
\end{equation}

\noindent with $A_{0}=1-v_{a}^{4}$. Thus, $M_{ADM}=$ $\sqrt{1-v_{a}^{4}}M$.
Probably an observer at any huge $r$ would see the mass $M_{ADM}$\ at the
center of the galaxy.\newline

There is a strong reason to consider with reserv the analysis near the
horizon: we are considering a time averaged photograph of the space-time,
which in the center would be strongly time dependent due to the presence of
the oscillaton, and the stationarity fails to be a good approximation, thus
it is not possible to know under the present approach which are the exact
roles played by the binding energy of the oscillaton, the negative energy of
the dust and the self-energy of the whole central system; such features
could be known by performing the evolution of Einstein's equations and in
fact represent a topic itself which will be discussed elsewhere \cite{futuro}%
. \newline

For this model, there is a contributor which does not fulfill the energy
conditions, this is the reason to call it ``toy model'',
although an observer at infinity will see the sum of all energy
densities of the components. In other words, the amount of matter, with
negative or positive energy density, is the quantity which determines the
value of $m(r)$, the ``balck hole mass'' concentrated at radius $r$\ of the
toy galaxy, not the contributors separately. Only if the observer could
measure the contributions of the energy density very close to the center of
the galaxy, it could recognize them. At infinity, the observer will only
measure $M_{ADM}$, i.e. it will see a Black-Hole-like metric at the center
of the galaxy which horizon lies at $r=2M_{ADM}$.\newline

We present in Figure 2 the situation when parameter $b$ takes different
values, playing thus the role of the core radius in the usual dark halo
hypothesis. Here we show that for an observer at infinity, the test
particles close to the galaxy center behave as if there were a black hole of
mass $M_{ADM}$ in the center. The event horizon avoids that this observer
could see inside of the horizon surface, the velocity of test particles is
higher for closer test particles and the rotation velocity decays as $1/%
\sqrt{r}$ in this region. For particles far away from the center, the
rotation curves behave just as the rotation curves given by the contribution
of the dark matter halo in a typical galaxy.\newline

\begin{figure*}
\centerline{ 
\epsfxsize=5cm \epsfbox{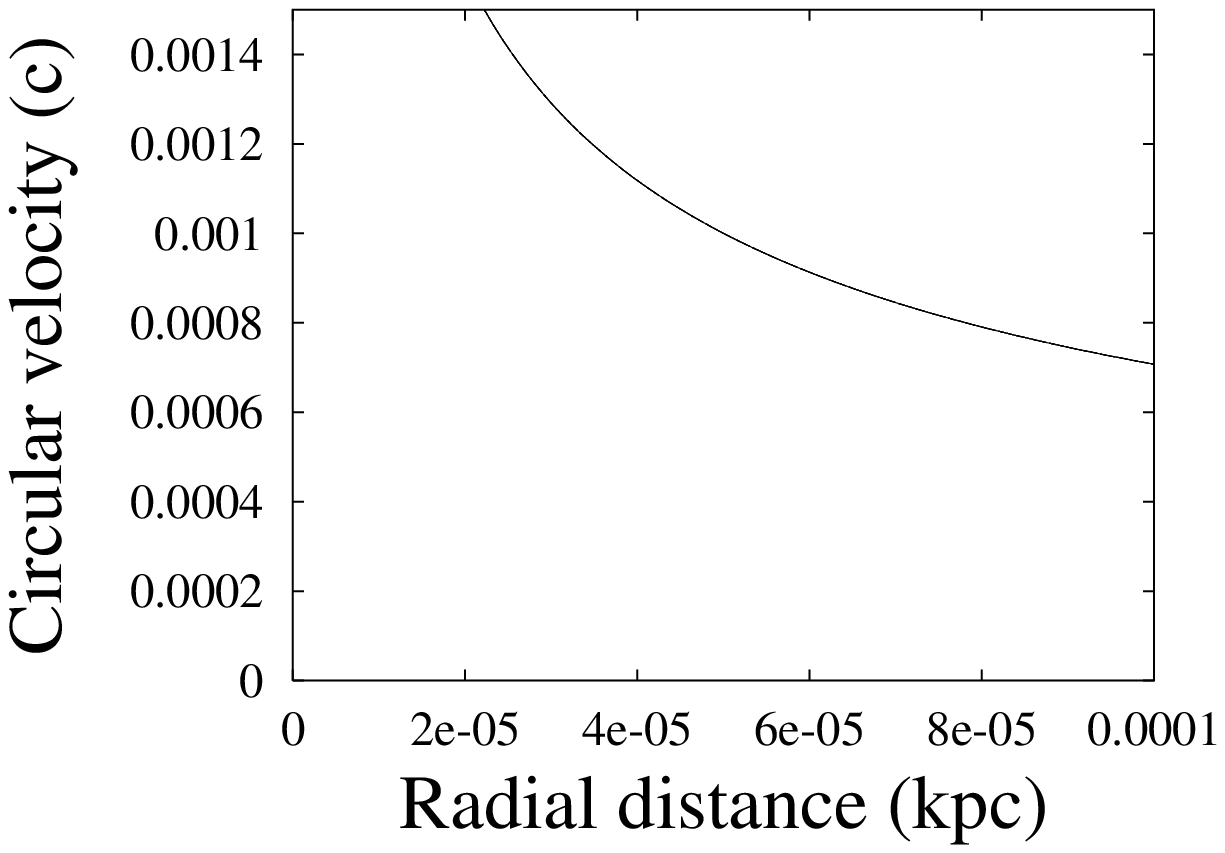} \epsfxsize=5cm \epsfbox{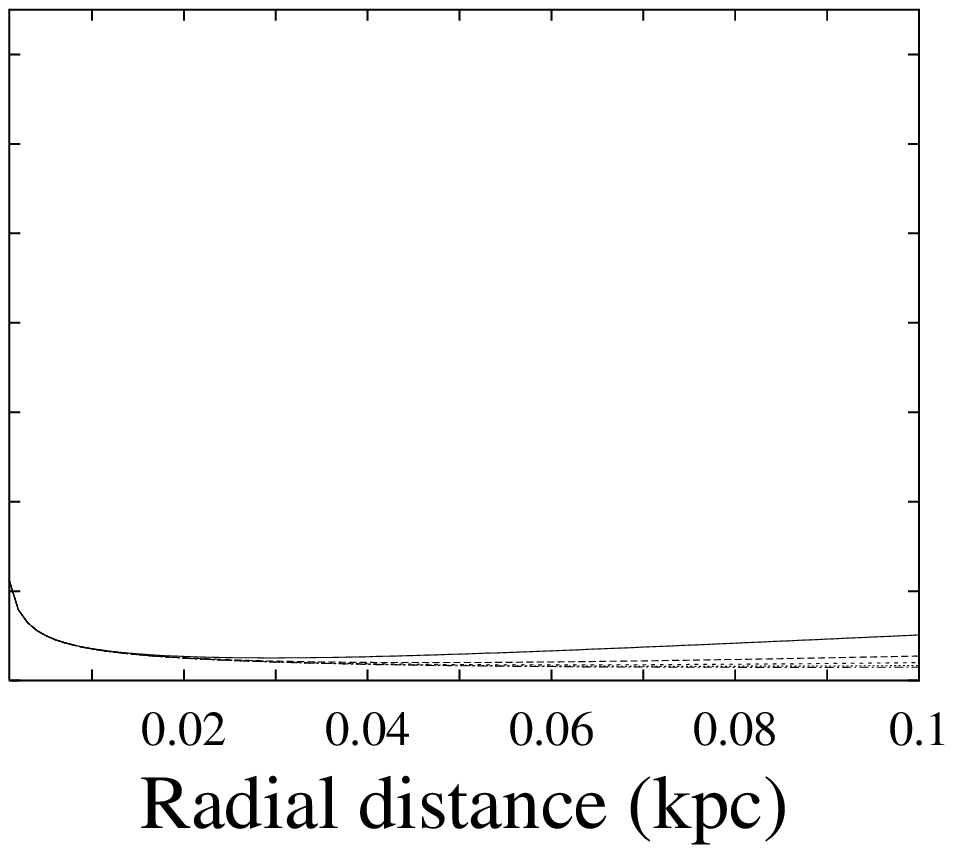}
\epsfxsize=5cm \epsfbox{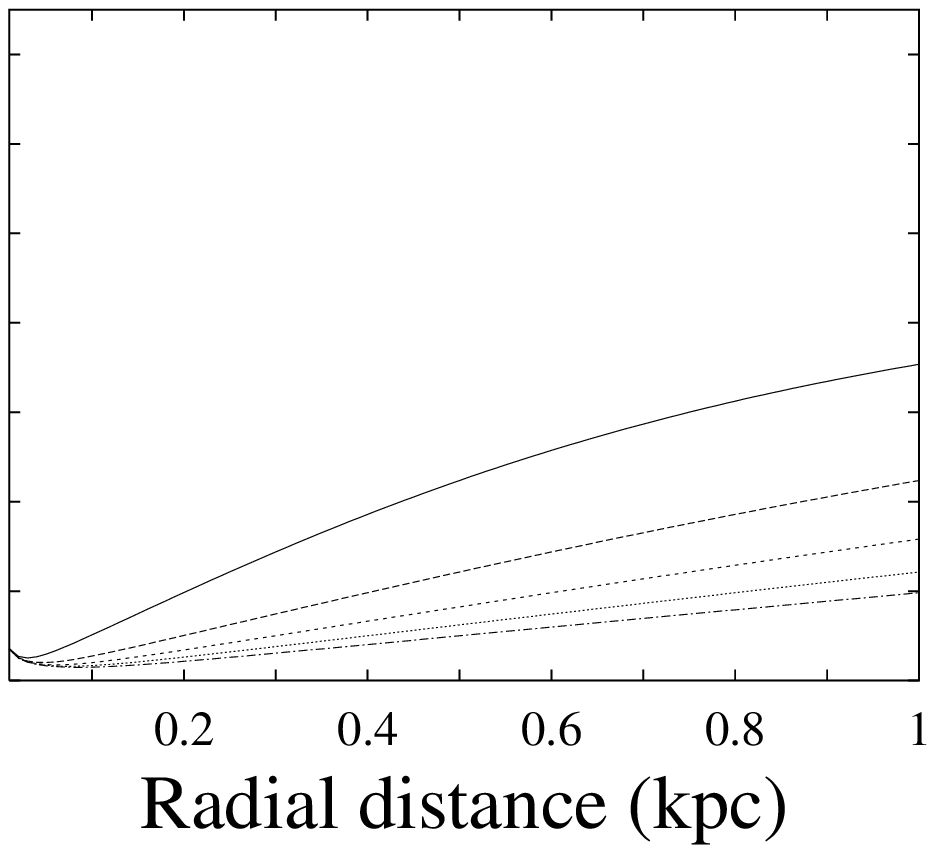} \epsfxsize=5cm \epsfbox{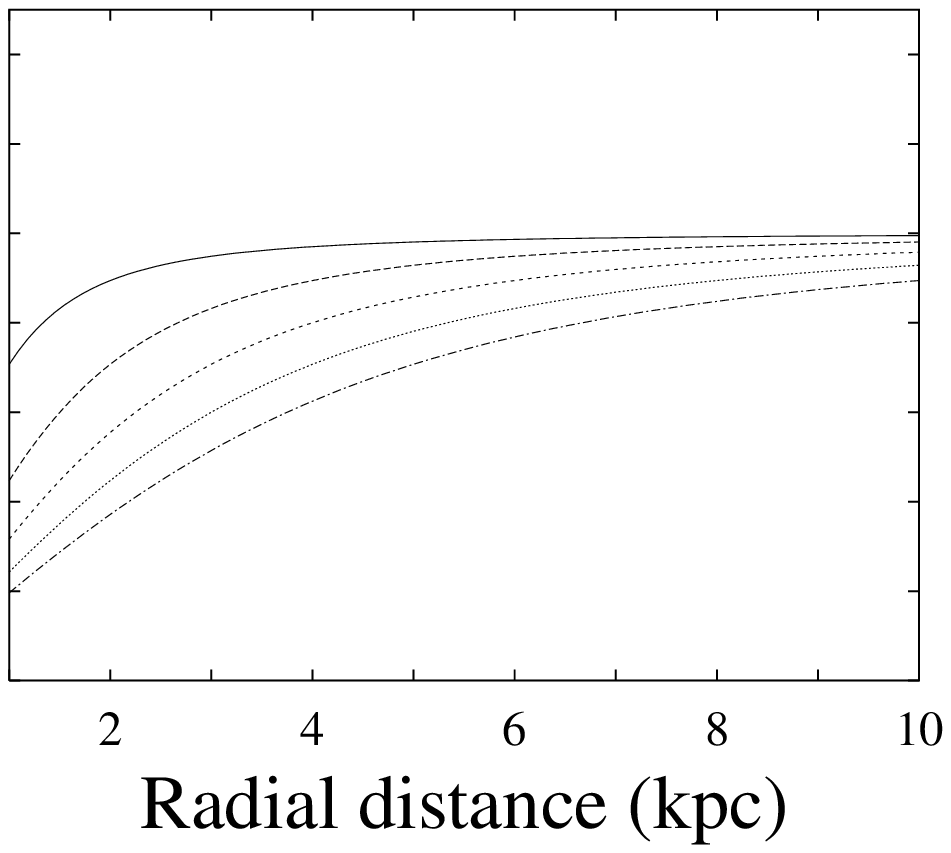}
}~\vspace{0.5cm}
\caption{We show the rotational curves when the parameter $b=1,2,3,4,5$kpc.
It is evident that in such case the shape of the curve associated with the
dark matter changes, but in the center remains unchanged.}
\label{fig:fig2}
\end{figure*}

In Figure 3 the fit of the curves is done using the observed rotation curves
of some dwarf galaxies, whose dark matter contribution is extremely
dominating and therefore are considered as the {\it test of fire} for a dark
matter model in galaxies. In general, for disc galaxies, the fit\ of the
rotation curves using this metric is analogous to that in reference
\cite{DMCQG}.
It seems then that metric (\ref{met_bosones}) is a good approximation for
some late stadium of the space-time of a spiral galaxy; it is a good
approximation of a ``long exposition photograph'' of a galaxy.\newline

\begin{figure*}
\centerline{ 
\epsfxsize=5cm \epsfbox{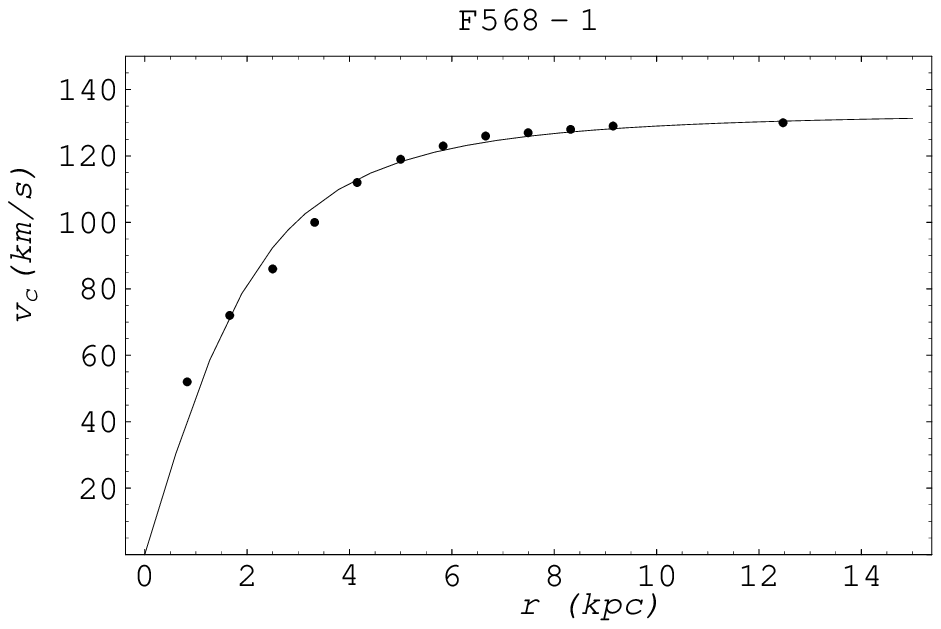} \epsfxsize=5cm \epsfbox{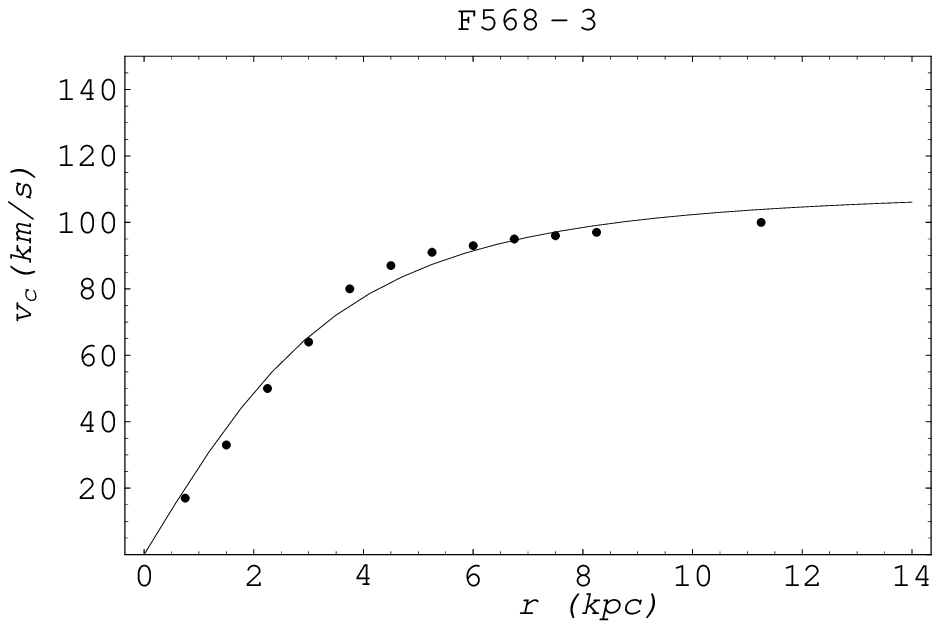}
\epsfxsize=5cm \epsfbox{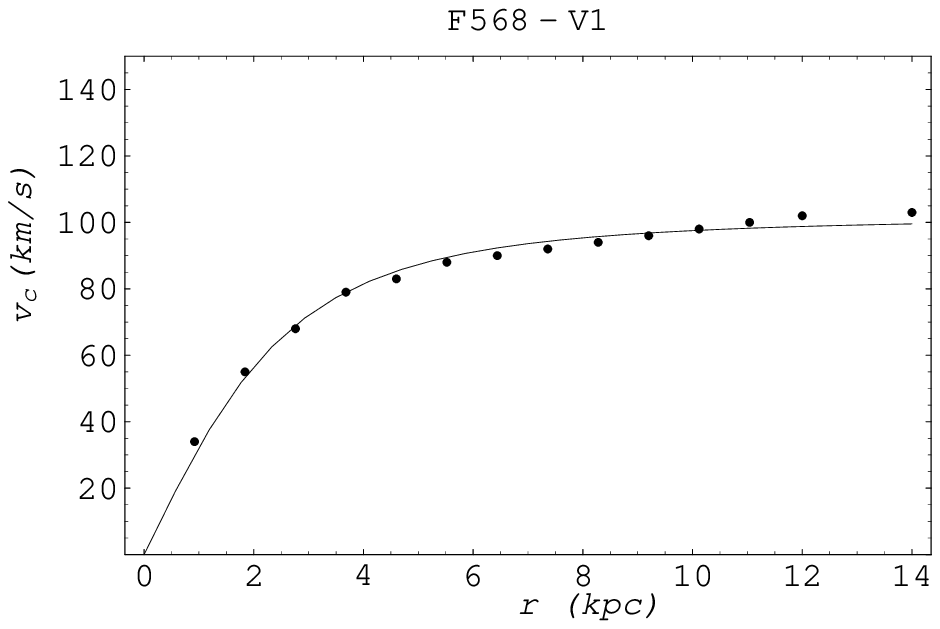}
}~\vspace{0.5cm}
\caption{Rotation curves of three dwarf galaxies are shown. The units are as
usual: kms${}^{-1}$ in the vertical axis and Kpc in the horizontal one.}
\label{fig:fig3}
\end{figure*}

\section{Discussion}

The configuration we found is as follows: in the innermost region of the toy
galaxy, it lies a time averaged oscillaton and dust, providing the picture
of an object very similar to a black hole for an observer far away.
Moreover, the resting scalar field should accommodate in the outer parts.
This picture is precisely very similar to the one required for a current
galaxy, which center would be dominated by a supermassive object and it
should posses a halo region dominated by the presence of dark matter. The
corresponding line element (\ref{met_bosones}) is free of naked
singularities and is not asymptotically flat. Instead, it behaves exactly as
a black hole near the coordinate center and it is asymptotically constructed
to provide the {\it flat curve condition} \cite{SPHPRD}.\newline

The main difference between this results and previous research supporting
the scalar field dark matter hypothesis in galaxies, is precisely that in
this case we have been able to extend the solution towards the galactic
core. We have shown two important statements: First, an averaged spherically
symmetric source field exists that supports a real scalar field under the
assumption of non asymptotic flatness. Second, a fully relativistic
treatment is able to explain the kinematic of test particles in the whole
background space-time of a galaxy at once.\newline

Therefore, we believe that the metric presented here could be a first
approximation of the galactic space-time provided the presence of coherent
field (a Bose-Einstein condensate) instead of usual matter. Even when our
solution accounts for negative energy densities of some components,
observers at infinity are not able to know it. Furthermore, for such
observers the test particles around the central object behave as if there
would be a black hole of mass $M_{ADM}$.

\acknowledgements{We would like to thank Luis Arturo Ure\~na and Dario Nu\~nez 
for many helpful discussions and remarks. 
This work is supported by CONACyT M\'exico under grant
34407-E.}\newline




\begin{references}
\bibitem{QSDMCQG}  T. Matos and L. A. Ure\~{n}a-L\'{o}pez, Class. Quantum
Grav. {\bf 17}, L75 (2000).

\bibitem{COSPRD}  T. Matos and L. A. Ure\~{n}a-L\'{o}pez, Phys. Rev. {\bf D
63}, 63506 (2001).

\bibitem{DMCQG}  F. S. Guzm\'{a}n and T. Matos, Class. Quantum Grav. {\bf 17}%
, L9 (2000).

\bibitem{SPHPRD}  T. Matos, F. S. Guzm\'{a}n and D. N\'{u}\~{n}ez, Phys.
Rev. D {\bf 62}, 061301 (2000).

\bibitem{Hu}  W. Hu, R. Barkana, and A. Gruzinov , Phys. Rev. Lett. {\bf 85}%
, 1158 (2000).

\bibitem{vladimir}  V. Avila-Reese, C. Firmani, A. Klypin and A. V.
Kravtsov, MNRAS {\bf 309}, 507 (1999).

\bibitem{CROSS}  T. Matos and L. A. Ure\~{n}a-L\'{o}pez, to be published.
E-print astro-ph/0010226.

\bibitem{apj}  L. Ferrarese and D. Merritt, ApJL, {\bf 539} L9 (2000). K.
Gebhardt {\it et al}, Astrophys. J. Lett. {\bf 539} L13 (2000).

\bibitem{diego}  D. F. Torres, S. Capozziello and G. Lambiase, Phys. Rev. 
{\bf D 62}, 104012 (2000).

\bibitem{seidel91}  E. Seidel and W. Suen, Phys. Rev. Lett. {\bf 66}, 1659
(1991).

\bibitem{seidel94}  E. Seidel and W. Suen, Phys. Rev. Lett. {\bf 72}, 2516
(1994).

\bibitem{luis}  L. A. Ure\~{n}a-L\'{o}pez, to be published. E-print
gr-qc/0104093.

\bibitem{tkachev}  A. Riotto and I. Tkachev, Phys. Lett. {\bf B 484}, 177
(2000).

\bibitem{jeremy}  J. Goodman, E-print astro-ph/0003018.

\bibitem{peebles}  P.J.E. Peebles, E-print astro-ph/0002495.

\bibitem{futuro}  M. Alcubierre, F. S. Guzm\'an, T. Matos, D. N\'u\~nez,
L. A. Ure\~na-L\'opez and P. Wiederhold, to be published.

\bibitem{ma}  T. Matos, Ann. Phys. (Leipzig) {\bf 46}, 462 (1989). T. Matos,
J. Math. Phys. {\bf 35}, 1302 (1994). T. Matos, Math. Notes {\bf 58}, 1178
(1995).
\end{references}
\end{document}